\documentclass[
  journal=bjps,
  year=2022,
  volume=2,
]{cup-journal}

\usepackage{graphicx} 

\usepackage{booktabs,microtype,siunitx,tabularx}
\title{Quantify how space mission influence geopolitical dynamics? A security and social policy approach.}

\author{Andrea Russo}
\affiliation{University of Catania, Italy}
\alsoaffiliation{University College Dublin, Ireland}
\email{Andrea.russo@phd.unict.it}

\author{Davide Coco}
\affiliation{University of Rome, Sapienza University, Italy }

\doi{--}

\received {--}
\revised  {--}
\accepted {--}
\published{--}


\usepackage{color}
\definecolor{darkgreen}{rgb}{0,0.5,0}
\definecolor{purple}{rgb}{1,0,1}
\definecolor{orange}{rgb}{1.0, 0.5, 0.0}
\newcommand{\kibitz}[2]{\ifnum\Comments=1\textcolor{#1}{#2}\fi}

\newcount\Comments  
 \usepackage{amsmath}

\usepackage[super]{nth}
\usepackage{csquotes}

\begin{document}

\begin{abstract} 



We present a computational method to quantify the geopolitical impact of a space mission, based on the national budget and data logs of previous mission,  and evidencing how even if some missions succeed, they can bring negative effect to the sponsored country. 
The objective of this research is to study how the success (or failure) of a space mission can bring an economical and political benefit (or loss) to a country.
By retrieving various data, including sentiment from \#hashtags related to the considered space missions, national budgets for space exploration, and the reliability of space launch systems, from social networks, public institutions, and online repositories, we propose an equation to evaluate the geopolitical importance of a space mission for a particular country or space agency.
The geopolitical equation can be used by public institutions or private companies to estimate the potential impact of a space mission on the public opinion and international relationships, which can be either positive or negative, as even successful missions may negatively affect the international relationships and negotiation with some countries and their partners.


Also we combine the ideology of classic social policy with a security and space mission point of view, to enlighten cultural, institutional, and political limits in public spending decisions. 

\end{abstract}

\section{Introduction}

----------------------------------





Since the end of World War II,
the proposal of ambitious space programs by governments and national space agencies has always been a means not only to push forward space exploration and research but also to alter the prestige and geopolitical influence in the international context. 
For instance, the social policy action by the \nth{35} United States president John Fitzgerald Kennedy to invest \$25 billion (1961 US dollar value) in the Apollo mission [\cite{bozzo_not_2018}] was not only a social investment to the increase public work with high qualification skills, but also a geopolitical plan action against the URSS space expansionism. \\
Geopolitical value appears since the first space missions, for instance, after Russian first space satellite "Sputnik 1" successfully orbited the Earth.
The Space Race that characterized the \nth{20} century, was actually a geopolitical and propaganda race to determine which country would have finally had access (and conquered) the \enquote{new and endless world above us}.
This geopolitical space race has been sustained by a huge effort from a social policy prospective. 
The \$25 billion USD for the Apollo mission on 12th September 1962 (same day of the \enquote{Address at Rice University on the Nation's Space Effort} speech by United States President John F. Kennedy to further inform the public about his plan to land a man on the Moon before 1970), are the equivalent of \$231 billion USD on 7th February 2022 [Appendix 1].\\
Nowadays, superpowers like the United States, China, or the Russian federation, have increased the frequency of space missions to show their presence and value in a geopolitical and international perspective.
The surge of space missions' proposals in the last decades was due also to the cost of access to space, which significantly decreased thanks to the development of reusable launch systems, performing hardware, and IT and IoT improvement.\\

Space missions have attracted huge money investments 
by public and private actors, with a social and business impact, 
due to the their potential economic return and their socioeconomic impact, as the design of a space mission encourages public high quality work and many public services originate from space activity (GPS, global mapping, high speed connection, global communication, and many others) [\cite{ASI2020}].

Given the aforementioned result, thus the spin-offs of space services to the population, can we define space missions (both research and security missions) as social policy, given the socio-economic effects and the infrastructure-related services?\\  
The security policy of the US government in early 2000s launched by the \enquote{Pentagon} (i.e., the headquarters building of the United States Department of Defense) highlights how the difference between social policy and security policy became less evident.
Indeed, the financial investment was aimed at developing a security program to defend social and public infrastructures and resulted in the X37-B program, a small shuttle that could defend  other satellites, which provide social services, from Russian and Chinese physical and cyber-attacks, thus interlacing the security and social aspects [\cite{X37B}].
Today, the relationship among social policy, security, space missions, and geopolitics is more intricate and complex than we could have imagined years ago.\\

Designing a space mission is an extremely difficult task, with a high probability of failure due to the complexity of aerospace systems and the harsh conditions under which they are supposed to operate.
The launch system plays an essential role in a space mission, as rockets must be \enquote{perfect} systems that respond seamlessly to all the perturbations that they experience during the atmospheric ascent and the exoatmospheric flight [\cite{benedikter2021convex,benedikter2022convex}].
Every phase of the ascent trajectory must be carefully studied and planned before flight, as the margin of error is extremely small, even for the apparently simple scenarios.
A first distinction by difficulty in space launches can be made by considering suborbital flights and orbital flights.
In the former, the greatest difficulty is reaching a high altitude and then re-enter the atmosphere before completing a full revolution around the Earth. 
Suborbital flights generally cross what is called the Karman line, an imaginary line at an altitude of 100 km (330'000 ft) above sea level that conventionally marks the boundary between the Earth's atmosphere and outer space. 
Although it is not a necessary requirement, the apogee (i.e., the maximum altitude) is a benchmark for more or less complex suborbital flights, as it indirectly determines the speed that the vehicle will reach during re-entry and therefore the thermal and mechanical stresses that it will undergo.
For orbital flights, instead, 
the inherent difficulty depends on the target orbit that the payload must be released into.
The range of possible scenarios increases enormously when it comes to orbital flights from one planet (or celestial body) to another. 
For instance, a flyby, that is, the short passage of a high-speed probe near a celestial body, is seen as an extremely critical moment compared to the simple time spent cruising.
On the other hand, the orbit insertion of a probe around a celestial body is a more critical moment than the flyby because it requires several active maneuvers that involve multiple simultaneously operating systems.
In a similar way, landing on a planet or asteroid is even a more critical accomplishment, as it involves much more complex operations.\\

This paper is inspired by the growing amount of usable data from social and space science. 
During recent years, social science has acquired methods and skills to collect data and use it like hard science to highlight and identify social patterns or social dynamics. 
Information technology has also increased the quality of social research, thanks to huge advancements in algorithms and data analysis.
This social-informatics improvement allows social science to connect with others disciplines, such as space science, engineering, and complex systems.
Social science can finally prove or evidence social complex interactions in political science and others social disciplines. 
For example, in this work, we try to evidence a complex interaction between space missions and international cooperation, based on risk-cost-benefit analysis and political affinity.

Due to the great inherent complexity of aerospace projects, international cooperation allows for mitigating the risks and costs (both financial and time-related) of space missions. 
There are several examples that show that the cooperation among national space agencies or research institutes has brought benefits to all the parties involved, not only relative to the economic return of scientific discoveries and patented technologies, but also to a positive outcome in terms of reputation and geopolitical prestige associated with these missions.

Besides the technical aspects, the organization and management of a space mission are also quite challenging because every political interaction or action during the mission has a series of \textit{emerging behaviors} in international affairs, and nonlinear interactions affect also reactions on political, economical, and security layers, which go beyond the space system [\cite{dittmer2014geopolitical}].
Complexity science tries to explain the dynamics of a system (e.g., physical, biological, social, or economical) that bring a different result than the expected one. 
A simple way to define a complex system is when the whole system is bigger than the sum of its parts.
Since space missions are related not only to social and security policies but also to international relationships and geopolitical dynamics, then they can be considered as complex systems.\\

Social policy, social network activity, and space exploration are just a slice of the whole part, but sufficient to understand how space missions influence the geopolitical strategy.
In complexity sciences, the online social network activity patterns consist of successive, spike-like perturbations, generated by consecutive shocks bearing conceptual similarities to the tremors preceding or following an earthquake [\cite{lymperopoulos2017dynamic}].
Indeed, the earthquake's dynamics follow the Self-Organized Critically (SOC) model [\cite{bak1991self}], present not only in nature but also in social systems [\cite{dmitriev2019self}]. \\
These dynamics could have been studied only with an hard science methodology and high quality data, which, in the pasts years, could have been very hard to collect.

In this paper, we used computational method to collect high quality data with a rigorous methodology to understand geopolitical and public policy effect from space mission. 
More precisely, we acquired data from social networks like Twitter, to evaluate the sentiment of space missions, which evidences the social reaction about the related space event. 
The sentiment analysis has been used by many others scientist over different research areas, and, if combined with a high quality computational method, it could highlight important patterns related to social events.
This methodology has been called Computational Social Science (CSS) [\cite{cioffi2014introduction}]. 

\subsection{Related Work}
Computational Social Science, since its introduction, 
provided a valuable approach to asses and explain social dynamical events. 
In this short section, we present some noteworthy works, relative to computational social complexity and sentiment analysis, helpful to understand our project.

Joseph Downing and Richard Dron [\cite{downing2020theorising}] have contributed to the understanding of constructivist security by analysing social media outputs to understand who is influential in the security debate.
Jie Yin et al. [\cite{yin2015using}] have focused on analyzing Twitter messages generated during humanitarian crises and disasters. 
They presented relevant methods for burst detection, tweet filtering and classification, online clustering, and geotagging to classify whether or not a tweet is talking about a disastrous event.
A disaster type classifier groups tweets according to representative disaster types: earth-quake, flooding, fire, storm, other disasters (e.g., traffic accident and civil disorders), and non-disaster.

Sancheng Peng et al. [\cite{peng2017social}] have worked on a framework that quantifies social influence in mobile social networks (Phones). 
The social influence of users was measured by analyzing the SMS/MMS-based communication behaviors among individuals. 
They have also revealed and characterized the social relations among mobile users through the analysis of the entropy of friend nodes and the entropy of interaction frequency. 
The extensive analytical results demonstrate that the influence spread of their proposed method outperforms a random method and a degree-based method.

Kolli et al. [\cite{kolli2017quantifying}] have quantified the predictability of cascade volumes in online social media communications, and tried to understand to what degree are future cascade trajectories predictable and to what degree are they random. 
The predictability analysis in their work reveals that for methods that combine information on frequency with temporal correlation of the trajectory, 
provides the theoretical limit on predictability. 
Hence, their methods such as AR and ARMA models that rely on the time order of data have the potential to achieve prediction accuracy as high as 83\% in
the MemeTracker dataset and 87\% in the Twitter Hashtag dataset. 
Correspondingly, these methods have the potential to achieve prediction accuracy as high as 94\% in the MemeTracker dataset.

Cihon and Yasseri [\cite{cihon2016biased}], have written a short review that considers a small number of selected papers on computational social science related to sociology and social science; they analyze their theoretical approaches, review their methodological innovations, and offer suggestions given the relevance of their results to political scientists and sociologists. 
They evidence how the sentiment analysis content using semantic and sentiment analytic algorithms on individual users opinions from the 15M movement in Spain ( analyzing up to 200 tweets on the topic per user) is a promising technique for future studies of political activity, and, indeed, any activity, on Twitter.

Vidgen and Yasseri [\cite{vidgen2020detecting}] built a multi-class classifier that distinguishes between non-Islamophobic, weak Islamophobic, and strong Islamophobic content. 
Accuracy is 77.6\% and balanced accuracy is 83\%. 
They applied the classifier to a dataset of \num{109488} tweets produced by far right Twitter accounts during 2017. 
While most tweets resulted not Islamophobic, weak Islamophobia was considerably more prevalent (\num{36963} tweets) than strong (\num{14895} tweets).

\subsection{Geopolitical Effects of Space Missions}

Sentiment analysis has been used in different research areas and with different goals. 
In this paper, we apply it to assess the geopolitical effect of space missions by collecting data from latest and legacy space mission that have or have not been successful.
Improving the success rate of space missions implies, from a geopolitical standpoint, an improvement of the international status.
However, on the other hand, a failure can damage the relationship among international partners. 
The success of the Apollo 11 mission by NASA made the United States the winner of the space race and raised its geopolitical value even though many milestones were reached earlier by the URSS (first orbiting satellite and first human in space, to name a couple).
Recently, numerous space missions were successfully launched and fulfilled their planned goals or even performed beyond expectations, receiving positive reception from the public opinion and altering (or consolidating) the international status of the involved countries. 
Noteworthy examples of recent successful missions are the Ingenuity helicopter, the first helicopter to fly outside the planet Earth, which was sent to Mars together with the Perseverance rover as part of NASA's Mars2020 mission, the Rosetta mission, which carried Philae, the first spacecraft ever to accomplish a soft landing on the surface of a comet (67P/Churyumov-Gerasimenko), launched in 2014 by ESA (European Space Agency), the latest James Webb Space Telescope, placed in a very challenging orbit (Sun–Earth Lagrangian point L2) in 2021 by NASA and ESA, and the Chang'e 4 mission, featuring the first soft landing rover on the far side of the Moon, lunched in 2018 by CNSA (China National Space Administration).

Even when a mission succeeds, there may be criticism from the society or even consequences and repercussions from others international actors, affecting the geopolitical status and international relationships of the sponsoring country.
A computational social science methodology [\cite{lazer2020computational}] 
 [\cite{lazer2009computational}] 
[\cite{cihon2016biased}] could assess how much space missions could increase or decrease (in case of failures or partially successful missions) the geopolitical status of each country.
In particular, the reaction of people on social networks generally gives a "feel strong" status to each country during international negotiations.
Also, a statistical qualitative evaluation of the success-to-failure ratio of space missions and rocket launches indicates the reliability of space launch systems and mission design and management of every country.
The geopolitical value depends on other people's perception of strength.
Therefore, if many people make negative comments (or have a negative sentiment) toward something that has been accomplished by some entity, implicitly it will come to change other people's perception of it, and thus also its geopolitical value.

Over the past decades, the goal of numerous space missions was
not only to meet the social policy agenda of each country or related to scientific investigations, but also to reinforce the strength of international relationships between countries.
For instance, the ExoMars mission has helped reinforcing the relationship between ESA and Roscosmos, the Russian space agency, before the Russian-Ukrainian war. 
Likewise, the DART mission is a cooperation between ASI (Italian Space Agency) and NASA, which strengthened the bond between European countries and the United States in space operations.
The number of cooperative space missions holds the promise to significantly increase in the near future, not only to strengthen and seal international relationships, but also to reduce the cost and time needed to design and accomplish a space mission.

The United States was the first country to understand this new political frontier [\cite{bozzo_not_2018}], and its actions have paved the way (from the industrial side) for the design of new engines and the world’s first reusable rocket, giving a huge advantage from the other space competitors.
The objective of this paper is to quantify the geopolitical score of each state, and evaluate how much it may fluctuate depending on space missions.

\subsection{Paper Outline}

This paper is organized as follows. Section 2 introduces the research hypotheses on space missions and geopolitical dynamics. Section 3 describes the methodological approach. 
In Section 4, we present the considered data sets and discuss the results. Section 5 explores the opportunity for a Space Mission Proposal by ESA, which can improve its geopolitical score.
A conclusion section ends the manuscript.


\section{Research hypotheses}
Our research hypothesis, is to establish whether it is possible to create an equation that gives a geopolitical scores inherent to space missions, since, we think that as a 'social policy', are included all those services deriving from space missions that also increase the security and geopolitical prestige of the state. 
To equate the outcome of space missions, as a public policy, is not so wrong given the increases in public services and highly skilled labour that are initiated at the beginning and end of each space mission.
Defence missions can also be part of an increase in work and services to citizens. Indeed, the US, Russian Federation, China, and EU countries use defense budgets to organize space missions for security and communication operation. 
Space missions can provide useful information to the state, companies and citizens, such as weather alerts to prevent catastrophes, prediction and prevention of potentially dangerous events, increased communication areas (internet and telephones) and also precise information on possible antagonists of the state to which they belong. 
In fact, over the past decades, many administrations have financed security space mission to obtain data about their competitors (using for example spy satellites) or just to amplify the satellite communication network. 
Anyway, the political tension about the dynamical and unpredictable security situation between US and China, did not touch only the military and economic sector, but have reached also the space and intelligence system [\cite{Giannellitensione}].
As a matter of fact, Admiral Rob Bauer, the Dutchman who since June 2021 chairs the NATO Military Committee, the highest military body of the Atlantic Alliance, did not turn around when it comes to describing the framework of the challenges pressing on the Euro-Atlantic area. 
He said that "Russia and China are increasing their respective military capabilities, conventional and nuclear, space and cyber: it's a fact, not an opinion." [\cite{StefanoNATO}]\\

However, space missions can also be used to improve conditions between states. 
Indeed, international events, such as space missions, are often used as a time to reconcile or strengthen relations between states. 
In fact, projects between the US and ESA are not only tools for providing jobs and services to citizens, thus moving the economy, but are also areas of cultural inclusion and knowledge, which is jealously guarded, especially the last one since it can be copied by foreign competitors. 
There can therefore be real international as well as economic strategies to get a space project off the ground. 

In order to calculate, how space missions have a geopolitical impact, hence on security and citizen services, we took a large part of the space missions inherent to the various countries.


\subsection{Considered Space Missions}

We searched which kind of recent (i.e., after social networks existence) space missions had had repercussions in the geopolitical and security domains, and we collected data for the missions reported in Table~ \ref{tab:1}.
The list includes only missions sponsored by countries that develop launch vehicles, as launch success and failure data are used as a way to estimate the country's experience in space missions (see Section~Methodology or Table \ref{tab:4}).

\begin{table}[htb!]
    \centering
    \begin{threeparttable}
    \caption{Space mission data collected}
    \label{tab:1}
    \begin{tabular}{l|c|c}
    \textbf{Mission} & \textbf{Country}  & \textbf{Result} \\
     Tianwen-1 & China & Succeeded\\
     Tianhe&China&Succeeded\\
     ASAT&Russia&Succeeded\\
     Mars 2020&US&Succeeded\\
     James Webb &US/EU&Succeeded\\
     Rosetta&EU&Semi-Succeeded
    \end{tabular}
\end{threeparttable}
\end{table}

The Tianwen-1 is a Chinese interplanetary mission to send a robotic spacecraft on Mars.
It was launched July 23, 2020 [\cite{China_tianwen}], and it landed on Mars on May 14, 2021 [\cite{China_tianwen2}]. 
The mission consisted of an orbiter, a lander, and a rover called "Zhurong".
The mission succeeded, and NASA's associate administrator Thomas Zurbuchen and the director general of Roscosmos Dmitry Rogozin congratulated the China National Space Administration (CNSA).

China has also lunched the Tianhe core module, which is the first module of the Tiangong space station, on April 29, 2021 atop a Long March 5B rocket [\cite{China_tianhe}]. 
The core stage of the LM-5B crashed back to Earth on Saturday, May 8, 2021, after 10 controversial days that captured the world's attention and started a wider conversation about orbital debris and the responsibility for the return of spent stages.\\
\\
On November 15, 2021, Russia conducted a direct-ascent anti-satellite test (ASAT), destroying one of its own space objects, a defunct satellite, in low-earth orbit [\cite{Russia_Asat}].
The test captured international attention and was quickly and widely condemned as threatening and irresponsible action, not least for the cloud of uncontrollable debris it created, which will endanger both space assets and human spaceflight for years to come.
Other countries in the past have already organised ASAT missions, like the Mission Shakti (India), the ASM-135 ASAT (US) and the 2007 Chinese anti-satellite missile test (China). 
Others to object in the wake of the test included Australia, the European Union, Japan, NATO, and South Korea. 
China and India -- the two countries other than Russia and the US that have previously conducted destructive ASAT tests —- are yet to comment publicly. 
Also, following the Russian ASAT mission, the International Space Station started emergency procedures due to the debris, closing its security hatches while the crew sheltered.\\
\\
Mars 2020 is a Mars rover mission [\cite{NASA_mars2020}] that includes the rover Perseverance and a small helicopter called Ingenuity.
It was launched from Earth on July 30, 2020 and landed in Martian crater Jezero on the 18th of February of the following year.
The Mars 2020 mission is forming part of NASA's Mars Exploration Program, which will continue with a sample return from Mars.
Ingenuity is a robotic helicopter that demonstrated the technology for rotor-craft flight in the extremely thin atmosphere of Mars, becoming the first controlled helicopter on another planet. 
The budget for the Perseverance rover was US\$2.8 billion in 2020 and was cheaper than its predecessor, the Curiosity rover, which costed \$3.2 billion. \cite{Curiosityrovercost}\\
\\
The James Webb Space Telescope [\cite{NASA_JWST}] is a space telescope designed primarily to conduct infrared astronomy. 
NASA led the development of the telescope in collaboration with ESA and CSA (Canadian Space Agency).
The mission duration is expected to be about 20 years and, the time planning for all the missions was 20 years (10 for planning and 10 for realization). 
It was Launched on December 25, 2021 by the contractor Arianespace from the Centre Spatial Guyanais, with an Ariane 5 rocket. 
The James Webb space telescope had a total budget of USD $\sim$ 9.70 billion (2002 to 2021) and has several scientific goals, including
the search for light coming from the first stars and galaxies that formed in the universe after the Big Bang, the study of the galaxy, star, and planet formation and evolution,
and studying planetary systems and the origins of life.\\
\\
The Rosetta mission [\cite{Rosetta}] was a space probe built by the ESA (European Space Agency) launched on March 2, 2004. The mission's goal was to perform a detailed and comprehensive study of comet 67P/Churyumov–Gerasimenko (67P). On August 2014, the spacecraft reached the comet and performed a series of manoeuvers to eventually orbit the comet at distances of 30 to 10 kilometres. 
The probe also housed a lander called Philae, which unfortunately was unable to last long on the comet's surface after a less-than-perfect landing.
This was, indeed, the first mission landing on a comet. 
Yet, despite the problems, the probe's instruments obtained the first images from a comet's surface, and several instruments made the first direct analysis of a comet, sending back data that would be analysed to determine the composition of the surface.
The mission cost was about $\sim$ 1.3 billion € (US $\sim$ \$ 1.8 billion).\\
\\
It is worth noting that even when the mission succeeds, it may happen that the mission goal has external or side effects that provoke a social or political reaction. 
Such reactions is spread, when there is something that does not belong to the common day-order, or that may provoke instability in some country-system.
We collect data from Twitter to see the social reaction to the space missions in Table~\ref{tab:1}. 
In all those cases, the missions succeeded, but, in some cases, the mission can cause a social reaction due to side effects. 

\section{Methodology}
Online social networks have evolved into valuable sources of information and pervasive communication platforms where people, businesses, and organizations generate and share content, build relationships, and join public conversations. In this online ecosystem, social networks are where information propagation is affected by external sources of influence, such as mass media, socioeconomic circumstances, advertising, or events, giving rise to collective intelligence or collective behaviour patterns.

We have collected social reactions from Twitter, for every mission in Table~\ref{tab:1} and compared them with the difficulty and quality of the national space organization and the country that launched the space mission to quantify the "Authority and the status of power" status in geopolitical interaction dynamics. 
To simplify this operation, we create an equation called GSS, to quantify how space mission can influence the geopolitical feelings during international affairs dynamics.  

The geopolitical space score (GSS) index is the geopolitical score value from each country, depending on the result of the spaceflight mission, related to statistical and social events [Equation~\eqref{eq1}]. 

\begin{equation} \label{eq1}
\begin{split}
GSS = \frac S{G*B} * (F+Q) \\
\end{split}
\end{equation}
\\ 
$S$ is the Sentiment value from the Twitter event; $B$ is the amount of  money  invested (budget); $G$ is a difficulty rate associated with the country that launched the space mission, "not because they are easy, but because they are hard" that imply Geopolitical effects; $F$ is related to the success or Failure of the mission; and $Q$ stands for the statistical Quality and difficulty of the country in spaceflight launch organization.    



The $S$ and $Q$ parameters are evaluated by data scientists through statistical methods. 
The $G$ score is the only parameter that needs a subjective value because it depends on personal evaluation and hypotheses to evaluate the difficulty rate of reaching that goal.

\subsection{S Factor}
To collect data for the selected topic, we use the public API by Twitter and Tweepy. 
Both services use permission from Twitter to obtain and gather data.
We collected a total of $\sim$ 7000 tweets, but any downloaded topic needs revisions and a cleaning process to increase the quality of the research. This has significantly reduced the volume of the tweets. 
We used the same methodology for each topic to obtain standard and quality data.
In addition, to obtain the correct amount of tweets 
for each day we use getdaytrends.com, a specific site where it is possible to monitor every topic in real time as well as aged topics.

To calculate the sentiment for the selected topic, we used the VADER sentiment analysis tools provided by MIT. 
The VADER sentiment quantifies each selected post or sentence's negative, neutral, or positive sentiment, giving in the end of the analysis, a compound score, i.e. the average between all sentence.   

However, to evaluate the sentiment for every mission, it is necessary that each mission be very important to the general public or, at least, sufficiently \textit{viral} among the space community. 

\subsection{G Factor}
As mentioned above, the $G$ factor is the only parameter without computational or statistical data. 
This parameter evaluates the country that launches the space mission, corresponding to a difficulty rate implying geopolitical effects. 
For example, if a small state succeeds in a mission with the same budget and other factors in comparison to a big state such as the US etc., the small state will get a bigger bonus. \\
Unfortunately, there is no universal value factor that gives a score to quantify the difficulty of space missions.

Geopolitics is a social evaluation, therefore deriving from the perceptual fluctuations of people. 
Since people make up a complex system, they cannot give a univocal value to the G factor, because it always depends on the value of the individual and on the oscillations (provoked by news, newspapers, friends, etc.) that modify the perception and evaluation of individuals and society on certain topics.

Due to this problem, we decided to self-evaluate the difficulty of a space mission, even if it is hard to make an assessment that takes into account everything. 
There are many pros and cons, and, in each case, we know that people cannot evaluate sufficiently well the difficulty of a space mission. 
Yet, we believe that people can understand whether it is a surprise if a very small country alone (like Ireland or Pakistan) can accomplish, for example, to build a Martian base, and the US could not do it. 
We estimate a self-score from 0.1 to 1, where 0.1 is the maximum and 1 is the minimum possible score. 
For example, a sub-orbital lunch mission could have a 1 score, an orbital mission could be a 0.8 score. The Tianwen-1 could be evaluated as 0.3, like the Martian Ingenuity helicopter on Mars. 
A full Moon base could be evaluated as 0.2 (or 0.1 if it is on Mars). 
All other space missions beyond the state-of-the-art technology level cannot be evaluated, because, we do not have sufficient information to be able to carry out the mission successfully, like a human base on the surface of Mercury, Titan or Europa, for example. 



\subsection{B Factor}
In the equation, we thought it was important to evaluate, then quantify, the level of resources invested versus the expected outcome (successful or failed mission). 
We thought this based on the logic "why should a mission cost a lot of money, while it is possible doing the same things while spending fewer resources?". 
To evaluate and quantify this factor, we collected data from the most important space agencies of each country.
Since the experience gained in the design and management of past missions can be exploited in newer missions,  
we chose not to use a specific budget for each mission in the equation, but, rather, the annual funds dedicated to space missions.
In this logic, it is hard to quantify how much money the oldest mission have helped (with knowledge, moneys, competence, technology) the newest space mission.  
Also, huge space missions like the JWST, or the Rosetta mission's budget, grows up during years. The budget for the mission is spread over years, and we cannot only take a single space mission budget, because there are also many other funded missions different from the JWST in the same year. 
Therefore, we thought that the Budget factor, imply also public opinion logic.  
Public opinion is usually skeptic about spending money for space mission, because they did not (unfortunately) see the huge policy investment on research, security and works employments, "Rockets don't run on cash".
And nowadays, is still difficult to quantify the investment return (knowledge, moneys, competence, technology) form the policy investment from space mission. 
In addition, the budget factor imply also a economic strength from the country that invest on space mission, 
For example, in the equation, if a small country achieve a successful space mission with a low budget, the GSS score will be higher than a the same country (or a bigger country) will achieve the same mission with a higher budget.\\
\\
However, to give an insight into space investment, in 2020, the policy plan for the major government in space mission amount of a \$73.98 billion, and it is the $\sim$ 0,927 \% as a share of gross domestic product (GDP), with a medium of $\sim$0,115875 \% for each country. 
The Organisation for Economic Co-operation and Development (OECD) as show the total value of space budgets from the G20 country [Table \ref{tab:2}], and we add in comparison, the years military budget for each country.

\begin{table}[htb!]
    \centering
    \caption{G20 government space and military budgets (2020)}
    \begin{threeparttable}
    \begin{tabular}{l|c|c|c}
    \textbf{Country} & \textbf{2020 Space Budget in Billions}  &\textbf{$\sim$ National Space Budget \%} &\textbf{$\sim$ National Military Budget \%}\\
     US&22.62&0.480\% & 3.74  \% \\
     Russia&3.58&0.210 \% & 4.26  \% \\
     France&4.04&0.122\% & 2.07 \% \\
     Japan &3.32& 0.076 \% & 1  \% \\
     Saudi Arabia &2.1 & 0.076 \% & 8.45  \% \\
     China&8.85&0.075\% & 1.75  \% \\
     Italy & 2.0 & 0.069 \% & 1.56 \% \\
     Germany & 2.40 &0.049  \% & 1.4   \%
    \end{tabular}
    \begin{tablenotes}[para]
  \item \% in billion U.S. dollars [Appendix 2]
\end{tablenotes}
\end{threeparttable}
\label{tab:2}
\end{table}
\subsection{F Factor}
The factor F, was needed to the equation to weigh/ponder the space mission, and it shows if the mission  succeeded or failed. The F factor is equal to 1 if the mission succeeded, or 0 if it failed.  

\subsection{Q Factor}
The factor Q, evidence the statistical risk factor and the reliability for each Country about spaceflight missions. Usually, every space mission have a risk factor, like the Apollo mission had the 95\% of failure [\cite{Nasarisk}], but this information (if they still made it) is not available to the public. 
Therefore, since we cannot obtain data for every single mission, we hypothesize a different evaluation method, so we rely as a risk factor on collecting data on the success/failure of each country's space launch. 
Those data give a specific statistical risk and reliability factor to each country, since year 2010. [Appendix 3]

It can happen that some space mission fail. In this scenario, we had imagined a failure factor that influence as a feature on GSS equation. 
The failures factor arises when you make mistakes over and over again, and in this case you always lose trust from others. The "Success/Authority improvement" comes from maintaining your own (high) standard for as long as possible.
We chose to not put the Failures Factor in the equation, because factor Q evidence all the space mission (satellite mission, supplying mission, scientific mission etc.), and not only the scientific space mission, like those we have chosen as subject of this paper (Table 1). 

A well know example for a Failures Factor, could be the Tianhe mission from China, that had unfortunately an uncontrolled stage reentry, and the vector crashed back to Earth without having the possibility to calculate the final crash site, due the amount of variables on the descent stage. 
Sadly, this inconvenience will arise often by the China, any time when they decide to add a core module on his space station, because the Long March 5B (Y2) cannot claim to get the core module in orbit (hooked to the Tianhe core stage), without losing control of the rocket on re-entry. [\cite{Chinarocket}]

The Long March 5B re-entry had provoke concern about the security for some city, because the rocket had an orbital inclination of 41.5 degrees, means the rocket body passes a little farther north than New York, Madrid and Beijing and as far south as southern Chile and Wellington, New Zealand, and could make it is reentry at any point within this area. With obvious concerns for those country. 

\section{Data \& Result} 


According to our model/equation, the most important and determinant score are the sentiment score, which refers as the reaction and evaluation about the space mission's result. The resulting online activity give us a social input valuable to quantify and qualify the international social reaction about space mission. 
Through the analysis of online activity topics, we identified the sentiment that describing the dynamic between space activity mission and geopolitical dynamics. 
In table \ref{tab:3} we show the sentiment results (S) from the most recently and most know space mission of the last decade.  
\begin{table}[htb!]
    \centering
    \begin{threeparttable}
    \begin{tabular}{l|c|c|c}
    \textbf{Mission} &\textbf{Hashtag} & \textbf{Sentiment}  & \textbf{Result} \\
Tianwen-1&Tianwen-1 &	0,46447	 & Succeeded \\
Tianhe*&ChineseRocket* & -0,05151 & Succeeded (Sides effects)\\
ASAT&ASAT & -0,16607 & Succeeded  (Sides effects)\\
Mars 2020*&Perseverance* & 0,428263 & Succeeded\\
Mars 2020&Mars2020* & 0,487525 & Succeeded\\
James Webb&JWST*&0,480994&Succeeded\\
Rosetta&Rosetta&0,429542&Semi-Succeeded
    \end{tabular}
    \caption{S factor - Sentiment analysis}
    \label{tab:3}
\begin{tablenotes}[para]
  \item [*] Actually, some hashtags are derived not from the mission itself, but from the effect achieved (losing control of the Chinese rocket on re-entry) or the main subject of the mission (perseverance). 
\end{tablenotes}
\end{threeparttable}
\end{table}

The results clearly evidence that an increase of side effect during mission, increases the negative score sentiments from the space mission community.

Regarding the economical resources invested (B) we collect data from different source, and arranged it on USD Billions dollars. Table \ref{tab:4} was the difficult one to make, because it is hard to obtain data from not-direct-democratic state, like China, and also because the inclusion of both civilian and military space budget for security missions. 
\begin{table}[htb!]
    \centering
     \caption{B factor - Space budgets}
    \begin{threeparttable}
    \begin{tabular}{l|c|c|c|c|c|c|c|c|c|c|c|c}
    \textbf{Country} &  \textbf{2010}  &\textbf{2011}  &\textbf{2012}  & \textbf{2013}  & \textbf{2014}  & \textbf{2015}  & \textbf{2016} & \textbf{2017} & \textbf{2018} & \textbf{2019}  & \textbf{2020} & \textbf{2021} \\
Japan &	1.67&	1.59&	1.68&	1.6&	1.76&	1.56&	1.33&	1.32&	3.06&	1.34&	3.32& 4.14\\	
Russia&	2.4	&3.8	& -&	5.6	&4.39&	2.42&	3.18& -	&	4.17&	3.58&	3.58&	1.92 \\
EU &4,19&	4,52&	4,56&	4,85&	4,85&	4,65&	5,95&	6,52&	6,35&	6,49&	5,52& 5,16\\
China	&-&-&-&-		&		2.66 &-	&	4.91 &-	&	5.83&	8.00&	8.85	&10.28\\
US &	18.72&	18.44 &	17.77&	16.86&	17.64&	18.01&	19.3&	19.50&	20.73&	21.5&	22.629&	23.27
    \end{tabular}
    \begin{tablenotes}[para]
  \item In billion U.S. dollars [Appendix 2] [Appendix 3]
\end{tablenotes}
\end{threeparttable}
\label{tab:4}
\end{table}

Therefore, the data from the statistical failures presence in our equation, that mark the quality of space launch system for each country (Q) are shows in Table \ref{tab:5}. 
The data are collected since the year 2010 to 2021 (the 2022 is not finished yet), show the total number of Core Stage Manufacture send in space, between overall launch log outside brackets and failures inside brackets. 
In the end, we shows also the total launch and failures between 2010 and 2021 and the failures percentage.
The failures percentage is the Q value in our equation, evidencing the failures probability to each space launch system-country for each launch.

\begin{table}[htb!]
\raggedright
\resizebox{0.85\textwidth}{!}{\begin{minipage}{\textwidth}
    \caption{Q factor - Statistical failures}
    \begin{threeparttable}
    \begin{tabular}{l|c|c|c|c|c|c|c|c|c|c|c|c|c|c|c}
    \textbf{Country} &\textbf{2010} &\textbf{2011} &\textbf{2012} &\textbf{2013} &\textbf{2014}  &\textbf{2015} &\textbf{2016} &\textbf{2017} &\textbf{2018}&\textbf{2019}& \textbf{2020}&  \textbf{2021} & \textbf{TOT} &\textbf{Failures \%} \\
     China&55(3)&	39(4)&	34(2)&	39(1)&	18(2)&	22(2)&	19(0)&	16(0)&	15(1)&	19(0)&	19(1)&	15(0)&	310 (16)&5,2 \\
     Russia&25(2)&	17(0)&	25(0)&	20(1)&	20(1)&	19(1)&	27(3)&	35(3)&	32(1)&	26(2)&	32(4)&	31(1)&	309(19)&6,1\\
     US&43(2)&	35(3)&	19(0)&	29(0)&	28(0)&	21(0)&	20(2)&	20(0)&	17(0)&	13(1)&	15(1)&	15(0)&	275(9)&3,3\\
     Europe&6(0)&	5(1)&	6(1)&	8(1)&	9(0)&	9(0)&	8(0)&	7(0)&	5(0)&	8(0)&	7(0)&	6(0)&	84(3)&3,6\\
     India&2(1)&	2(0)&	6(0)&	7(0)&	5(1)&	7(0)&	5(0)&	4(0)&	3(0)&	2(0)&	3(0)&	3(2)&	49(4)&8,2\\
     Japan&3(0)&	4(0)&	2(0)&	6(0)&	7(1)&	4(0)&	4(0)&	4(0)&	3(0)&	2(0)&	3(0)&	2(0)&	42(1)&2,4\\
     New Zeal.&6(1)&	7(1)&	6(0)&	3(0)&	1(1)&	-&	-&	-&	-&	-&	-&	-&	23(3)&13,6\\
    Iran&1(1)&	2(1)&	2(2)&	-&	-&	-&	1(0)&	-&	-&	3(2)&	1(0)&	-&	10(6)&60\\
    \end{tabular}
\end{threeparttable}
\label{tab:5}
\end{minipage} }
\begin{tablenotes}[para]
  \item Total launch by year (total launch failures by year) [Appendix 4]
\end{tablenotes}
\end{table}

\begin{figure}[hbt!]
    \centering
    \includegraphics{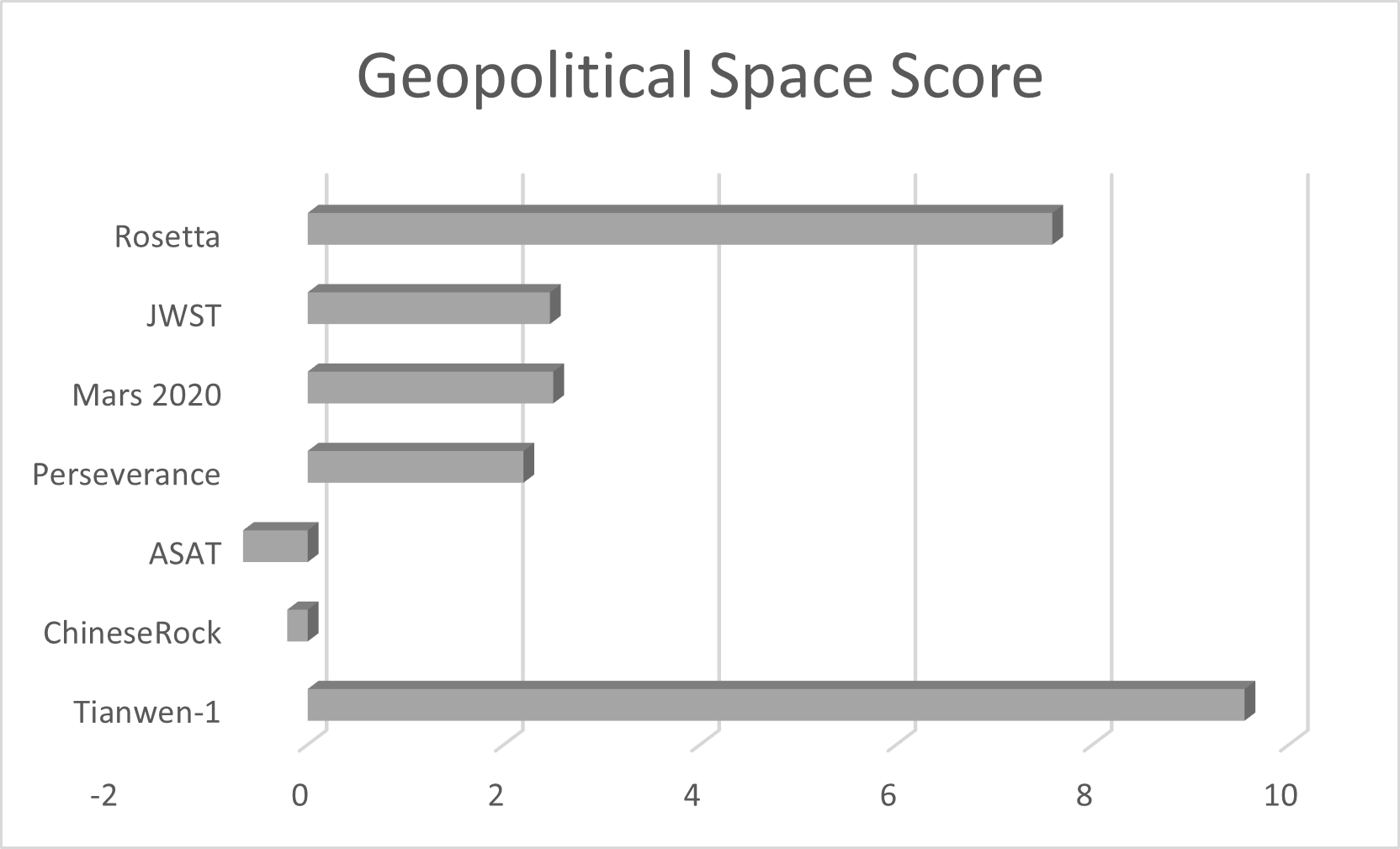}
    \caption{GSS's info-graphic of mains space mission}
    \label{fig1} 
\end{figure}

It is possible to see that the Chinese government have many more launch than the US and Europe, this is due to the low presence of space satellite (for communication and security mission) orbiting the Earth by the Chinese government. The Chinese government had invested a huge amount of money to get enough satellites to make public infrastructures work.  
Also, Russian launch operation have been increased since the 2011, due to the retirement of the space shuttle (last mission 21th July 2011), and as result, the US  astronauts had to traveling by Russian Soyuz spacecraft to get to the international space station.
We have compared the data from the equation (S, T, R and Q parameters), and in Figure \ref{fig1} and Table \ref{tab:6} we show the score after the mission succeeded or failed.

\begin{table}[htb!]
    \centering
    \caption{Geopolitical Space Score}
    \begin{threeparttable}
    \begin{tabular}{l|c}
    \textbf{Mission} & \textbf{Geopolitical Space Score}\\
     Tianwen-1 &	9,547438889 \\
ChineseRock	 & -0,208492857 \\
ASAT &	-0,659007937\\
Perseverance &	2,198416733\\
Mars 2020 &	2,502628333\\
JWST &	2,469102533\\
Rosetta	 &7,588575333
    \end{tabular}
 \begin{tablenotes}[para]
  \item                         
\end{tablenotes}
\end{threeparttable}
\label{tab:6}
\end{table}

As is easiest to see, there is a negative score (we have highlight it with red on Figure 1), due to the risk of the ASAT and the space debris rocket had on population on heart and on the ISS. 
Tianwen-1 and Rosetta mission have a high GSS score also because it was a first huge milestone for the respectively space agency (ESA and CNSA).
These data evidence the difference between a successful mission, achieving its goal; wile a successful mission, still achieving is goal but with side effect.

\subsection{GSS trend in time}
We have tried to quantify a GSS value for space Research Mission only between 2010 and 2021, for each country, due the fact that ASAT is a military space mission. So we have gather data from all factor for the equation, but unfortunately we did not get all the Sentiment Factor (S), because usually unknown space mission does not have much reaction on social network; As said before to evaluate the sentiment for each missions, those missions should be very important for the Humankind, or at least "virality" for the space community, and many mission unfortunately become known to major society only from newspaper and occasionally from TV news. 
To solve this problem, we estimate a medium 0,425 S factor from successful mission, and a medium -0,125 for failed missions. For the other well know mission (Table 1) we have used the original sentiment.
Regarding the economical resources invested (B) toward the data accumulated, we did not get the full budget planning over years for China and Russia, so we estimate a progressive linear regression between the missing data on table 4.
\begin{figure}[ht!]
    \centering
    \includegraphics [width=\linewidth] {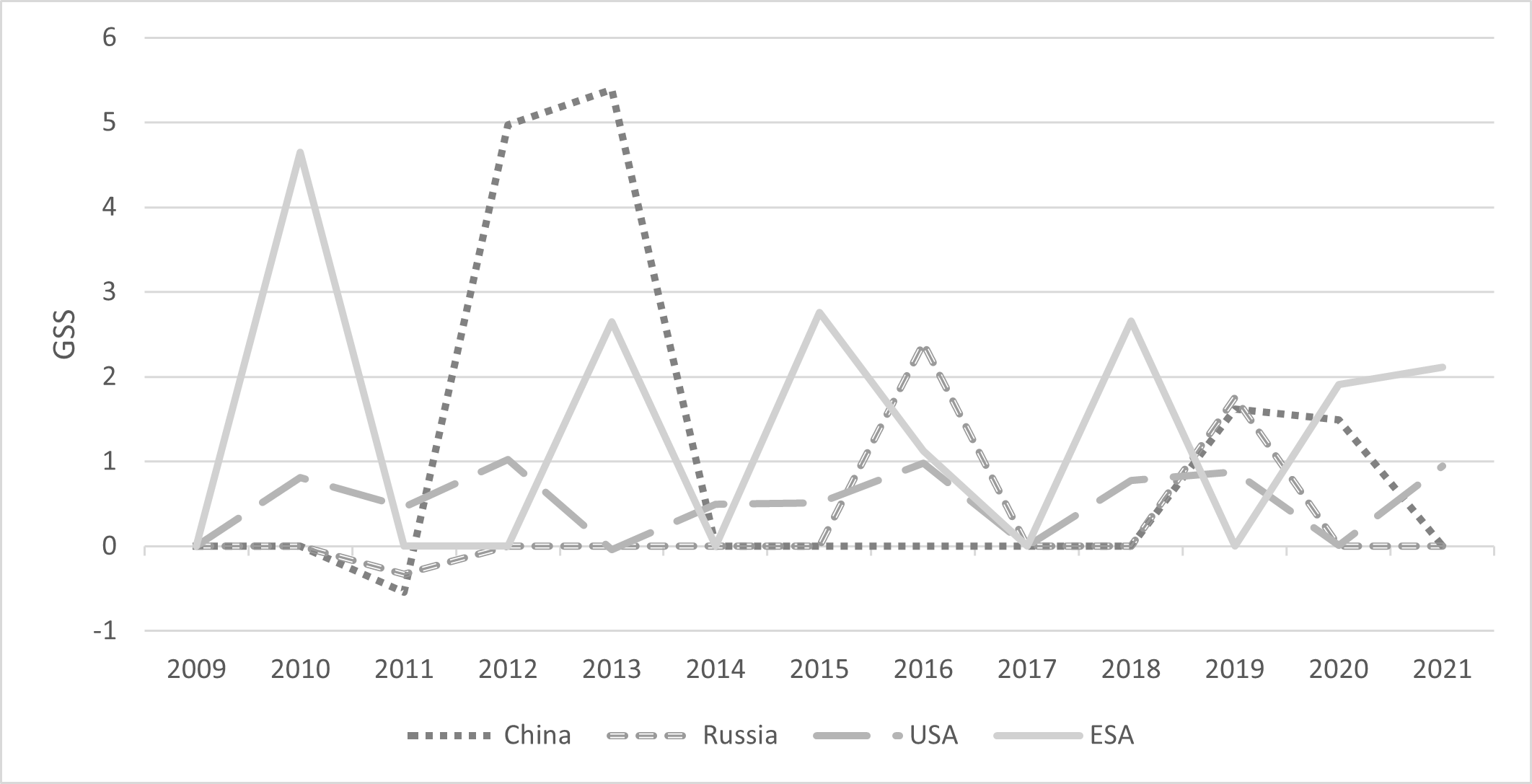}
    \caption{GSS - Research missions}
    \label{fig2}
  \begin{tablenotes}[para]
  \item Appendix 5 - 6
\end{tablenotes}
\end{figure}

As Figure \ref{fig2} shows, it is possible to see the fluctuations characterised by the missions. 
In addition, it is also possible to see the type of strategy and activity for each country/agencies. 

For example, ESA, which is characterised by the limitation of not having a launch station on the east coast, has specialised in international collaboration, and it is possible to  see how ESA manages missions with fluctuations of about one year (given the many collaborations, especially with NASA). 
Moreover, it should be noted that the Rosetta mission, like the Tianwen-1 mission, could be designed as a "baptism of independence", derived from "not because they are easy, but because they are hard" philosophy; Thy was indeed, the first to do that, showing his competence and skill upon the others superpowers space country like US and Russia. 
China, on the other hand, is concentrating on a few space research missions, since it is a new superpower and has yet to stabilise its infrastructure and network telecommunication on space.
Instead, the US have/had many active missions for a long time, in fact the score is relatively lower than the others precisely because of their resilience. People expect them to fail less than the others, so the astonishing/sensational felling can only be found in very difficult missions, or when they failed badly.

\begin{figure}[ht!]
    \centering
    \includegraphics [width=\linewidth] {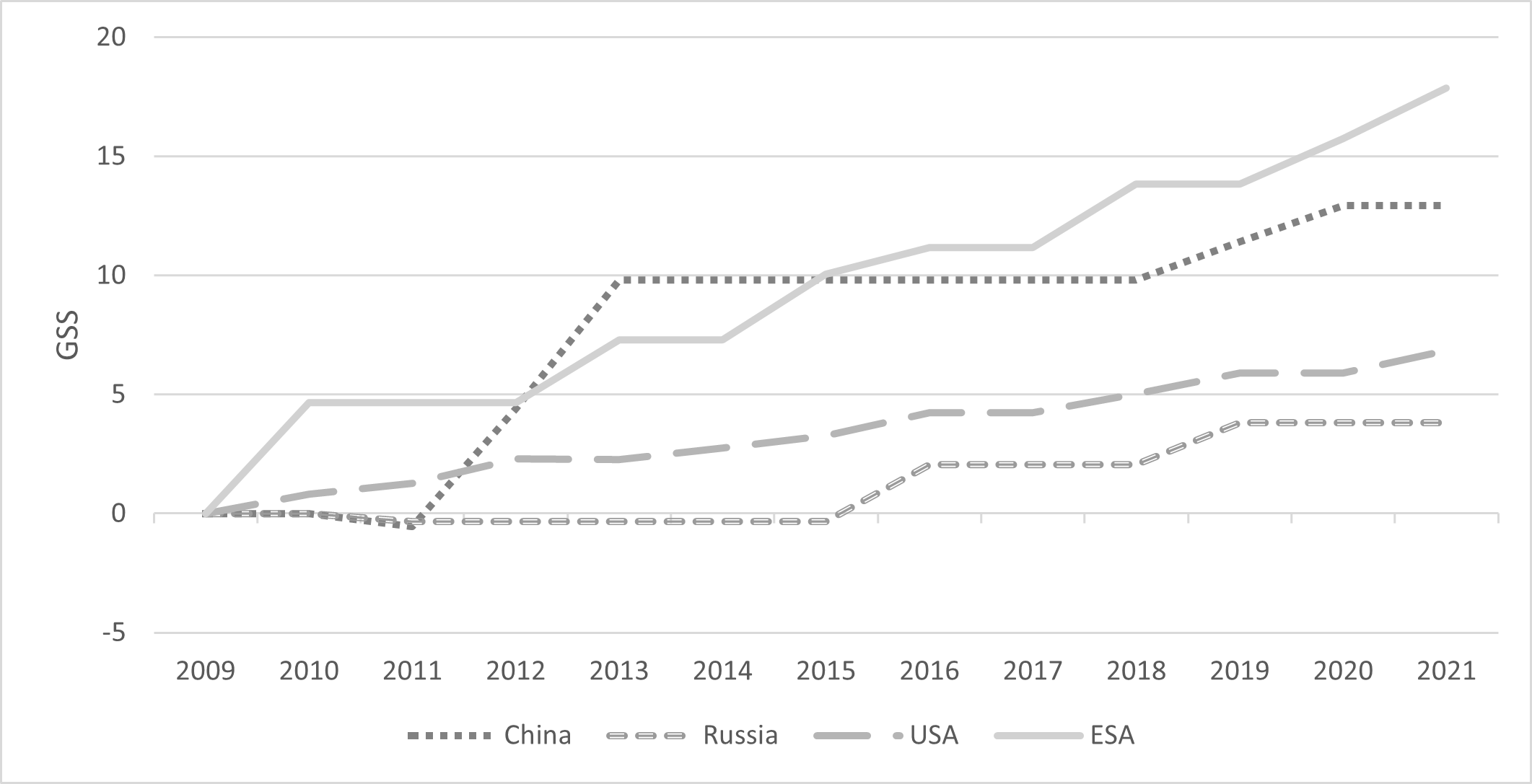}
    \caption{GSS - Research missions cumulative}
    \label{fig3}
  \begin{tablenotes}[para]
  \item Appendix 5 - 6
\end{tablenotes}
\end{figure}

\section{Analysis}
The analyses support the expectation that: 1) that it is possible to numerically assess geopolitical factors; 2) that even in the case of a successful mission, there is the possibility of a negative geopolitical feedback; 3) that public spending on space missions can also positively improve the geopolitical level of the investing state; 4) that the geopolitical level can vary over time, depending on the success of the missions, proportional to the investment. 

An increase in the social policy inherent in space missions, in addition to providing public employment with low-high qualification (and therefore removing people from the poverty line), would improve the risk factor (Q Factor) decreasing it to each future space mission, thus improving the geopolitical sentiment and also the geopolitical weight of the state.

In fact, as already mentioned, the most resilient state is the US. They are not first on the geopolitical level despite having a high number of missions both quantitatively and qualitatively, in indeed, this score is given by the fact that they rarely miss, thus showing a high quality work, given the quality of the workers.

\section{Conclusions}
Our study shows that can be possible using computational social method to quantify geopolitical dynamics for every country. 
In our case we have shown how space mission influence geopolitical dynamics, supported by a security and social policy approach. 
In a more specific way, we have demonstrated (1) A socio-physical method for assessing the geopolitical level of any country, based on its goal and its system-organization; (2) That space missions, even if successful, can bring negative sentiment, which goes to reflect on the geopolitical value to the state itself; (3) That social policy is also an investment in the defense and security system, and that increased investment of government spending on space missions, also has a spillover effect on geopolitical value.  
From a sociological and socio-physical point of view, through the contribution of a Computational social science model, the GSS equation can be used to evaluate the geopolitical level not only in the spatial domain, but also in other "competitions", such as sports competitions (Olympics, world sports championships) or music competitions or international wars, where there are enough social reactions (S Factor) and enough statistical data (Q - B Factor) to evaluate both the event and the organization/country participating in the event.  
Therefore, our work would emphasise the most general Political policy, and in the most specific way, the Social policy,  as not only a economical financial aids, but also a more complex systems, related to complex dynamics. 
As we demonstrated that Social policy is also funding the defense, security, and geopolitical systems. It has become not only an aid to the population, but much more. 
Moreover, an increase in the spending policy for space missions (which we define as part of social policy), also has a geopolitical impact on the state that promotes them, and by reflection, to its society. 

\section{Declaration of interest statement}
The authors declare no conflict of interest.

\section{Authors details}
\subsection{Andrea Russo}
Is a PhD candidate in Complex Systems at the University of Catania. 
He is currently working at the Department of Physics and Astronomy. He collaborated with CNR Ibam, he also has worked purely on projects involving technology and society. 
His main research field and interests are focused on the study and the development of Computational social method to explain social complexity, in particular field like Politics - Economics - Business and Defense-Security sector applications. 
ORCID: 0000-0003-3816-0539\\
Corresponding author. Email: Andrea.russo@phd.unict.it\\

\subsection{Davide Coco}
M.Sc. in Space and Astronautical Engineering at University of Rome "Sapienza", with several years of experience in mission design and preliminary studies of space missions.\\
His main goal is to further specialize and be involved in conceptual studies, system requirements definition, mission proposal writing, space system modeling and simulation, launcher or mission design and operations.\\
He is currently leading the design and development of Cubesat's subsystems in a small Italian startup, Human4Research.\\
ORCID: 0000-0001-8010-9468\\
Email: davide.coco@outlook.it

\begin{acknowledgement}
We thank Taha Yasseri for for his crucial help in the equation. 
\end{acknowledgement}







\printbibliography

\appendix

\section{Equivalence of the 1962 USD to 2022 USD}

More specifically: 231 408 697 113.64 USD \$ .\\
Inflation over the period: 825.63 \% \\
Index used: USCPI31011913 (Bureau of Labor Statistics).  \\
Initial Index: 309.12, Final Index: 2 861.33\\
Link=https://fxtop.com/it/conversione-valute-passato.php

\section{Space annual budget ($\sim$ approximately) over years and percentage of the national budget}

It is very hard to collect data for the most important Space Agency, this is due to the different money value (Yen - EURO - USD), and also the difficult to obtain data from not-democratic state, like China, due to information security and also due to the inclusion of both civilian and military space spending global security. \\
\\
Source:
\begin{enumerate}
    \item https://www.statista.com/statistics/745717/global-governmental-spending-on-space-programs-leading-countries/
    \item https://www.euroconsult-ec.com/press-release/government-space-budgets-driven-by-space-exploration-and-militarization-hit-record-92-billion-investment-in-2021-despite-covid-with-1-trillion-forecast-over-the-decade/\
    \item https://spacenews.com/op-ed-global-government-space-budgets-continues-multiyear-rebound/
    \item https://stacker.com/stories/2524/countries-spend-most-space-exploration 
    \item https://www.oecd.org/sti/inno/space-forum/space-economy-for-people-planet-and-prosperity.pdf
    \item https://global,jaxa,jp/about/transition/index,html
    \item esa.int
    \item https://en.wikipedia.org/wiki/Budget\_of\_NASA
    \item https://global.jaxa.jp/
\end{enumerate}
[\cite{spacebudget}]

Military budget: 
\begin{enumerate}
    \item https://databank.worldbank.org/reports.aspx?source=2\&type=metadata\&series=MS.MIL.XPND.GD.ZS\#
    \item https://www.defensenews.com/global/2021/04/26/the-world-spent-almost-2-trillion-on-defense-in-2020/
\end{enumerate}

\section{Billion U.S. dollars}

Exchange rate 1.1269 at 27/02/2022 02:50 28 Feb 2020 - 25 Feb 2022

\section{Statistical failures}
Annual space reports:\\
From 2010 to 2021 Launch Log\\
Source = https://www.spacelaunchreport.com/index.html

\section{GSS - Research Missions}

\begin{table}[htb!]
\raggedright
\resizebox{0.6\textwidth}{!}{\begin{minipage}{\textwidth}
    \caption{GSS related to space research missions}
    \begin{threeparttable}
    \begin{tabular}{l|c|c|c|c|c|c|c|c|c|c|c|c|c|c}
    \textbf{Country} &\textbf{2009} &\textbf{2010} &\textbf{2011} &\textbf{2012} &\textbf{2013} &\textbf{2014}  &\textbf{2015} &\textbf{2016} &\textbf{2017} &\textbf{2018}&\textbf{2019}& \textbf{2020}&  \textbf{2021} \\
  China	&	-	&	-	&	-0,541666667	&	4,969230769	&	5,383333333	&	-	&	-	&	-	&	-	&	-	&	1,615	&	1,490577956	&	-	\\
Russia	&	-	&	-	&	-0,334429825	&	-	&	-	&	-	&	-	&	2,389937107	&	-	&	-	&	1,75719055	&	-	&	-	\\
USA	&	-	&	0,808621288	&	0,464438894	&	1,026104163	&	-0,040508492	&	0,492783694	&	0,512710274	&	0,976252159	&	-	&	0,776134433	&	0,876356589	&	0,011896792	&	0,942333613	\\
ESA	&	-	&	4,647391567	&	-	&	-	&	2,648339061	&	-	&	2,762246117	&	1,128012708	&	-	&	2,654855643	&	-	&	1,903820817	&	2,11289354	\\
Japan	&	-	&	4,542027057	&	-	&	-2,976190476	&	-	&	4,277597403	&	-	&	-	&	-	&	5,740740741	&	-7,462686567	&	-	&	-	\\
TOT	&	-	&	3,332679971	&	-0,137219199	&	1,006381485	&	2,663721301	&	2,385190548	&	1,637478196	&	1,498067325	&	-	&	3,057243606	&	-0,803534857	&	1,135431855	&	1,527613577	

    \end{tabular}
\end{threeparttable}
\end{minipage} }
\begin{tablenotes}[para]
  \item GSS - Research missions
\end{tablenotes}
\end{table}

\section{Country Succeeded and Failed Space Research Mission}
\begin{table}[htb!]
    \raggedright
\resizebox{0.75\textwidth}{!}{\begin{minipage}{\textwidth}
    \caption{Country Succeeded(Failed) Space Research Mission}
    \begin{threeparttable}
    \begin{tabular}{l|c|c|c|c|c|c|}
    \textbf{Year} &\textbf{China} &\textbf{Russia} &\textbf{USA} &\textbf{ESA} &\textbf{Japan} &\textbf{India}  \\
2010	&	-	&	-	&	Deep Impact	&	Rosetta	&	Akatsuki	&	-	\\
	&	-	&	-	&	Stardust	&	-	&	IKAROS (Shin'en)	&	-	\\
2011	&	(Yinghuo-1)	&	(Fobos-Grunt)	&	Dawn	&	-	&	-	&	-	\\
2012	&	Chang'e 2	&	-	&	MSL Curiosity	&	-	&	(PROCYON)	&	-	\\
2013	&	Chang'e 3	&	-	&	(Deep Impact)	&	Gaia	&	-	&	-	\\
2014	&	-	&	-	&	MAVEN	&	-	&	Shin'en 2	&	Mangalyaan	\\
2015	&	-	&	-	&	DSCOVR	&	LISA Pathfinder	&	-	&	-	\\
	&	-	&	-	&	New Horizons	&	-	&	-	&	-	\\
	&	-	&	-	&	Dawn	&	-	&	-	&	-	\\
2016	&	-	&	ExoMars 2016 (Schiaparelli EDM lander)	&	Juno	&	ExoMars 2016 (Schiaparelli EDM lander)	&	-	&	-	\\
2017	&	-	&	-	&	-	&	-	&	-	&	-	\\
2018	&	-	&	-	&	Parker Solar Probe	&	MASCOT	&	Hayabusa2	&	-	\\
	&	-	&	-	&	MarCO A "WALL-E"	&	BepiColombo	&	BepiColombo	&	-	\\
	&	-	&	-	&	MarCO B "EVE"	&	-	&	-	&	-	\\
	&	-	&	-	&	OSIRIS-REx	&	-	&	-	&	-	\\
	&	-	&	-	&	InSight	&	-	&	-	&	-	\\
2019	&	Chang'e 4	&	Spektr-RG	&	New Horizons	&	-	&	(Minerva II-2)	&	-	\\
	&	-	&	-	&	Spektr-RG	&	-	&	-	&	-	\\
2020	&	Chang'e 5	&	-	&	Mars 2020	&	Solar Orbiter	&	-	&	-	\\
	&	Tianwen-1	&	-	&	-	&	-	&	-	&	-	\\
	&	Beidou	&	-	&	-	&	-	&	-	&	-	\\
2021	&	-	&	-	&	James Webb 	&	James Webb 	&	-	&	-	

    \end{tabular}
\end{threeparttable}
\end{minipage} }
\begin{tablenotes}[para]
  \item GSS - Research missions
\end{tablenotes}
\end{table}

\end{document}